\documentclass[10pt, A4]{article}
\usepackage[scaled=.89]{helvet}

\usepackage{textcomp}
\usepackage{pifont}
\usepackage[T1]{fontenc}

\usepackage{graphicx}

\usepackage{color}
\usepackage{url}
\usepackage[colorlinks=true, pdfstartview=FitV, linkcolor=black, citecolor=black, urlcolor=blue]{hyperref}

\usepackage{pgf,pgfarrows,pgfnodes,pgfautomata,pgfheaps,pgfshade}
\usepackage[english]{babel}
\usepackage{tikz}
\usepackage[overlay, absolute]{textpos}

\usepackage{verbatim}
\usetikzlibrary{arrows,shapes}
\usepackage[latin1]{inputenc}
\usetikzlibrary{trees,snakes}

\tikzstyle{format} = [circle, draw, thin, minimum width=22pt, fill=yellow!60]
\tikzstyle{dia} = [circle, draw, thin, minimum height=16pt, minimum width=22pt, fill=yellow!60]
\tikzstyle{format2} = [ellipse, draw, thin]
\tikzstyle{circ} = [draw, thin, fill=green!60, minimum height=18pt, minimum width=18pt]
\tikzstyle{medium} = [ellipse, draw, thin, fill=blue!30, minimum height=2.5em]
\tikzstyle{medium2} = [ellipse, draw, thin, fill=red!30, minimum height=2.5em]

\newcommand\timearrow{\begin{tikzpicture}[thick]
\coordinate
    child[grow=down, level distance=95pt]{
     node {{\small \emph{Past}}}
        edge from parent [arr0] 
    }
        child   [grow=north] {
            node {{\small \emph{Future}}}
            edge from parent [arr1] 
        };
\end{tikzpicture}}

\newcommand\arrowonline[1]{%
    node
    [pos=#1,sloped,anchor=center,draw=none,fill=none,shape=rectangle,inner
    sep=0pt,outer sep=0pt] {\tikz\draw[->,blue, dash pattern=on 0pt off 2pt]
    (-1pt,0pt) -- (0pt,0pt);}
}

\renewcommand\arrowonline[1]{}

\def\electronedge{edge from parent [electron] \arrowonline{0.55}}
\tikzstyle{photon} = [segment amplitude=0.7pt, segment length=5pt, snake=snake, draw=blue]
\tikzstyle{gluon} = [segment amplitude=4pt, segment length=5pt,
                     snake=coil, draw=magenta]
\tikzstyle{electron} = [draw=blue,->]
\tikzstyle{electronz} = [draw=blue,-<]
\tikzstyle{arr1} = [draw=black!40,->]
\tikzstyle{arr0} = [draw=black!40]
\tikzstyle{electron-} = [draw=blue,-<]
\tikzstyle{electron-z} = [draw=blue,->]
\tikzstyle{electron0} = [draw=blue]
\tikzstyle{electron1} = [draw=white]

\tikzstyle{level 1}=[level distance=1.5cm]
\tikzstyle{level 2}=[level distance=1.3cm, sibling distance=2.2cm]
\tikzstyle{level 3}=[level distance=1.3cm,sibling distance=1.6cm]

\newcommand{\helsink}[7]{

\begin{tikzpicture}[thick]
\coordinate
    child[grow=down]{
     node[dia] {$#7$}
        edge from parent [electron-] node [left=1pt] {$$}
    }
    child[grow=up, level distance=0pt] {
   child  {
        child   [grow=north] {
             node[circ] {$#2$}
            \electronedge
        }
          child [grow=south east]{ 
            node[format] {$#6$}
            edge from parent [electron-] node [right=1pt] {$$}
        }
       edge from parent [photon]
        node [right=1pt] {$#5$}
    }
    child {
         child [grow=south west]{ 
            node[format] {$#3$}
            edge from parent [electron-] node [left=1pt] {$$}
        }
       child  [grow=north]{ 
            node[circ] {$#1$}
            edge from parent [electron] node [left=1pt] {$$}
        }
                     edge from parent [photon] 
        node [left=2pt] {$#4$}
    }
};
\end{tikzpicture}}

\newcommand{\helsinkRETRO}[7]{
\begin{tikzpicture}[thick]
\coordinate
    child[grow=down]{
     node[dia] {$#7$}
        edge from parent [electron-] node [left=1pt] {$$}
    }
    child[grow=up, level distance=0pt] {
   child  {
        child   [grow=north] {
             node[circ] {$#2$}
            \electronedge
        }
          child [grow=south east]{ 
            node[format] {$#6$}
            edge from parent [electron-] node [right=1pt] {$$}
        }
       edge from parent [photon]
        node [right=1pt] {$#5$}
    }
    child {
         child [grow=south west, level distance=17pt]{ 
            node[format] {$#3$}
            edge from parent [electron-] node [right=1pt] {$$}
        }
       child  [grow=north]{ 
            node[circ] {$#1$}
            edge from parent [electron] node [left=1pt] {$$}
        }
                     edge from parent [photon] 
        node [right=2.5pt] {$#4$}
    }
};
\end{tikzpicture}}

\newcommand{\helsinkBIG}[7]{
\begin{tikzpicture}[thick]
\coordinate
    child[grow=down]{
     node[dia] {$#7$}
        edge from parent [electron-] node [left=1pt] {$$}
    }
    child[grow=up, level distance=0pt] {
   child  {
        child   [grow=north] {
             node[circ] {$#2$}
            \electronedge
        }
          child [grow=south east, level distance=50pt]{ 
            node[format] {$#6$}
            edge from parent [electron-] node [right=1pt] {$$}
        }
       edge from parent [photon]
        node [right=1pt] {#5}
    }
    child {
         child [grow=south west, level distance=50pt]{ 
            node[format] {$#3$}
            edge from parent [electron-] node [left=1pt] {$$}
        }
       child  [grow=north]{ 
            node[circ] {$#1$}
            edge from parent [electron] node [left=1pt] {$$}
        }
                     edge from parent [photon] 
        node [left=2pt] {#4}
    }
};
\end{tikzpicture}}

\newcommand{\helsinkz}[7]{
\begin{tikzpicture}[thick]
\coordinate
    child[grow=down]{
     node[circ] {$#7$}
        edge from parent [electron-z] node [left=6pt] {$$}
    }
    child[grow=up, level distance=0pt] {
   child  {
        child   [grow=north] {
             node[dia] {$#2$}
            edge from parent [electronz] 
        }
          child [grow=south east]{ 
            node[circ] {$#6$}
            edge from parent [electron-z] node [right=5pt] {$$}
        }
       edge from parent [photon]
        node [right=1pt] {$#5$}
    }
    child {
         child [grow=south west]{ 
            node[circ] {$#3$}
            edge from parent [electron-z] node [left=4pt] {$$}
        }
       child  [grow=north]{ 
            node[dia] {$#1$}
            edge from parent [electronz] node [left=4pt] {$$}
        }
                     edge from parent [photon] 
        node [left=2pt] {$#4$}
    }
};
\end{tikzpicture}}

\newcommand{\helABBa}{\helsink{C}{C}{B}{A}{A}{B}{A}}

\newcommand{\helABCa}{\helsink{C}{B}{B}{A}{A}{C}{A}}

\definecolor{ZurichBlue}{rgb}{.255,.41,.884} 
\definecolor{ZurichRed}{rgb}{1, 0, 0} 
\definecolor{ZurichGreen}{rgb}{.196,.804,.196} 
\definecolor{ZurichYellow}{rgb}{1,.648,0} 

\begin{document}
\title{Toy Models for Retrocausality\thanks{This note is based on a talk given at workshops at the University of Sydney and at Griffith University, Brisbane, in November, 2007. The slides for the Griffith University version of the talk are available online here:  \href{http://www.usyd.edu.au/time/price/preprints/RetroTalkGriffithNov07.pdf}{\url{http://www.usyd.edu.au/time/price/preprints/RetroTalkGriffithNov07.pdf}}. I am grateful to John Cusbert, Pete Evans, Eric \mbox{Cavalcanti} and other participants in those workshops  for helpful feedback, and to Steve Weinstein and especially Ken Wharton, for much helpful discussion since then. I am also indebted to the Australian Research Council and the University of Sydney, for research support.
}}

\author{Huw Price\thanks{Centre for Time, University of Sydney; email: \href{mailto:huw@mail.usyd.edu.au}{\url{huw@mail.usyd.edu.au}}.}}
\date{February 22, 2008}

\maketitle
\section{Motivation}
A number of writers have been attracted to the idea that some of the puzzling features of quantum mechanics might be manifestations of `reverse' or `retro' causality, at a level underlying that of the usual quantum description. The main motivation
 for this view stems from EPR/Bell phenomena, where it offers two virtues. First, as was noted by its earliest proponent,\footnote{The view was first proposed by Olivier Costa de Beauregard (1911--2007), a student of Louis de Broglie, whose first publication on the subject is (Costa de Beauregard 1953). Prof.~Costa de Beauregard (2005) reported that he had proposed the idea several years earlier, in 1947, but that de Broglie had forbidden him to publish it, until Feynman's work on the positron gave some respectability to the idea of ``things going backwards in time''. For more recent retrocausal proposals, see the references in (Sutherland 2006).} it has the potential to provide a timelike decomposition of the nonlocal
  correlations revealed in EPR cases -- i.e., as we would now put it,\footnote{The qualification is necessary because Costa de Beauregard's version of the proposal pre-dates Bell's
   work by more than a decade, of course.}  for the violation of the Bell inequalities in the quantum world. Second, Bell's derivation of his famous inequality depends explicitly on the assumption that hidden states do \emph{not} depend on future measurement settings -- so that its violation simply \emph{invites} a retrocausal explanation, at least from the point of view of anyone who has already been bitten by the retrocausal bug. 

Most people working in the foundations of quantum mechanics remain resolutely unbitten, however. It is common for the retrocausal option to be ignored altogether, or, as in this rather careful recent survey article, relegated to the footnotes with other \emph{unmemorabilia:} 
\begin{quote}
To be scrupulous, there are perhaps four other ways [i.e., other than nonlocality] that the correlations in [an EPR-Bohm] experiment could 
be explained away.  (1) One could simply `refuse to consider the correlations mysterious'.  (2) One 
could deny that the experimenters have free will to choose the settings of their measurement devices 
at random, as required for a statistically valid Bell-experiment.  (3) \emph{One could entertain the idea 
of backward-in-time causation.} (4) One could conclude that ordinary (Boolean) logic is not valid in our Universe.  I do not consider these escape routes because they seem to undercut the core assumptions necessary to undertake scientific experiments. (Wiseman 2005, my emphasis)
\end{quote}

What can a fan  of quantum retrocausality do at this point, to try to bring the proposal out of the footnotes and onto the main page? Well, there are two obvious strategies. The first is to construct explicit theories and models of quantum phenomena, embodying retrocausal principles. Various proposals of this kind are in the literature.\footnote{See, e.g., (Sutherland 2006) and the references therein, and (Wharton 2007).} The second  is to explore the conceptual foundations of the proposal -- e.g., to examine the basis of our ordinary causal intuitions, in order, perhaps, to uncover some deep-seated errors in reasoning, underlying intuitive objections to retrocausality. Again, some work of this kind is in the literature -- see, e.g.,  (Price 1996).

This note introduces a third strategy, which offers a promising complement to the other two, in my view. This third strategy aims to investigate retrocausality in general -- and hopefully, eventually, \emph{quantum} retrocausality -- by developing simple `toy models', to explain and elucidate its characteristics, and to explore its potential and peculiarities.

\subsection{Playing with models}

Much of the inspiration for this project comes from (Spekkens 2004), who proposes and investigates a `toy theory', as he calls it, to explore the issue as to which of the distinctive features of quantum mechanics might be explained by the hypothesis that the quantum description is `epistemic', rather than `ontic'. By analogy, one aspect of my third strategy -- admittedly, one that I make almost no progress with in this note -- is to use toy models to explore the question as to what quantum-like phenomena retrocausality might in principle explain.
The other aspect is to use such models as intuition pumps, or teaching aids, for clarifying and motivating the unfamiliar ideas involved in retrocausal proposals -- e.g., for looking for latitude in what Wiseman (\emph{op. cit.}) termed `the core assumptions necessary to undertake scientific experiments.'

The toy model described here shows how something that `looks like' retrocausality can emerge from global constraints on a very simple system of `interactions', when the system in question is given a natural  interpretation in the light of familiar assumptions about experimental intervention and observation. It yields nothing of a distinctively quantum nature, except a crude form of nonlocality. I present it in the hope that it may turn out to be a stepping stone to something more interesting, and especially in the hope that it will help to explain \emph{what the game is,} to people who still find retrocausality as unattractive a response to the quantum puzzles as fatalism, non-Boolean logic, or a shrug of the shoulders.

\section[The Helsinki model]{The Helsinki model\footnote{When I first presented this model at a workshop in Sydney in 2007, I explained that I had two reasons for calling it the Helsinki model: (i) I thought of retrocausality in QM as an elegant rival  to Copenhagen (though somewhere in the same neighbourhood, in some respects); and (ii) the model itself first occurred to me while I was stuck between flights at Helsinki airport, a couple of months previously. Contrary to an interjection at that point by my student, John Cusbert, the name has nothing to do with ``control by the Finnish state.''}}

The Helsinki model is defined by the following ingredients and principles:
\begin{itemize}
\item There are two kinds of primitive nodes, each the inverse of the other under reflection around the horizontal axis, and each comprising a meeting-point of three edges. If we interpret the edges as `particle world-lines', then the nodes represent two kinds of primitive `interaction': `pair production' and `pair annihilation'. (See Fig.~1)
\item Each edge has one of three `flavours', $A, B$ or $C$.
\item Each node must be strictly \textbf{inhomogeneous} -- i.e., comprising three edges of \emph{different} flavours -- or  strictly \textbf{homogeneous} (three edges of the \emph{same} flavour).
\item Pair production and pair annihilation must alternate, when the primitive nodes are linked together.
\item \emph{Successive} homogeneous nodes are prohibited. (See Figs.~2 \& 3)
\end{itemize}

\begin{figure}[h]
\vspace{0mm}
\centering
\mbox{\begin{tikzpicture}[thick]
\coordinate
    child[grow=down, level distance=44pt]{
     node {\mbox{\vrule width 0pt depth 0pt height  15pt {\small \emph{`Pair production'}}}}
        edge from parent [electron0] node [left=1pt] {$A$}
    }
    child[grow=up, level distance=0pt] {
   child[grow=north east]  {   
       edge from parent [electron0]
        node [right=2pt] {$B$}
    }
    child[grow=north west]   {
                     edge from parent [electron0] 
        node [left=4pt] {$C$}
    }
};
\end{tikzpicture}}\hspace{10mm}
\raisebox{0.8mm}{\mbox{\begin{tikzpicture}[thick]
\coordinate
     child[grow=down, level distance=44pt]   {node  {{\small \emph{`Pair annihilation'}}}
                  edge from parent [electron1] 
}
    child[grow=up, level distance=40pt]{node{}
 edge from parent [electron0] node [left=1pt] {$A$}
    }
   child[grow=south east]  {   
       edge from parent [electron0]
        node [right=2pt] {$C$}
    }
    child[grow=south west]   {
                     edge from parent [electron0] 
        node [left=3pt] {$B$}
 };
\end{tikzpicture}}}
\caption
[Figure 1]
{The two basic `interactions'.}
\end{figure}
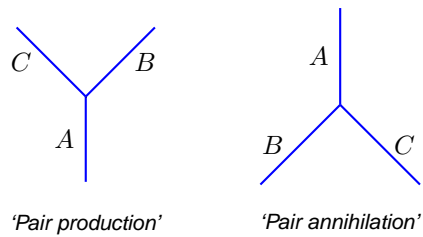

\subsection{Adding `time evolution', `preparation' and `observation'}

The bare dynamics of this model is `up-down' symmetric --  or \emph{time}-symmetric,  if we treat up-down as a temporal axis (Fig.~4). Given such a temporal interpretation, however, then it is very natural to imagine we can \emph{control} the inputs and \emph{read off} the outputs, as in Fig. 5. Here yellow circles represent `interventions', or `preparations' (values we can `choose to assign'); green squares represent `observations' -- values we simply `read off';  and the wavy lines represent the `hidden' sectors, that we can't directly control or observe. Note that the  two pair annihilations in Fig.~5 provide `measurements' of the hidden sectors, in the sense that if we know one input and the output, the rules uniquely determine the value of the second (`hidden') input.

\begin{figure}[t]
\vspace{0mm}
\centering
\begin{tikzpicture}[thick]
\coordinate
    child[grow=down]{
        edge from parent [electron0] node [left=1pt] {$A$}
    }
    child[grow=up, level distance=0pt] {
   child  {
        child   [grow=north] {
            edge from parent [electron0] node [right=1pt] {$B$}
        }
          child [grow=south east]{ 
            edge from parent [electron0] node [right=2pt] {$C$}
        }
       edge from parent [electron0]
        node [right=1pt] {$A$}
    }
    child {
         child [grow=south west]{ 
            edge from parent [electron0] node [left=2pt] {$A$}
        }
       child  [grow=north]{ 
            edge from parent [electron0] node [left=1pt] {$A$}
        }
                     edge from parent [electron0] 
        node [left=2pt] {$A$}
    }
};
\end{tikzpicture}
\caption
[Figure 2]
{\textbf{Disallowed} -- repeated homogeneous nodes.}
\end{figure}
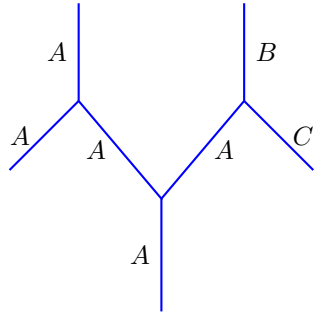
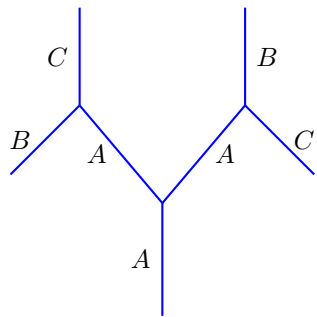
\begin{figure}[h]
\vspace{2mm}
\centering
\begin{tikzpicture}[thick]
\coordinate
    child[grow=down]{
        edge from parent [electron0] node [left=1pt] {$A$}
    }
    child[grow=up, level distance=0pt] {
   child  {
        child   [grow=north] {
            edge from parent [electron0] node [right=1pt] {$B$}
        }
          child [grow=south east]{ 
            edge from parent [electron0] node [right=2pt] {$C$}
        }
       edge from parent [electron0]
        node [right=1pt] {$A$}
    }
    child {
         child [grow=south west]{ 
            edge from parent [electron0] node [left=2pt] {$B$}
        }
       child  [grow=north]{ 
            edge from parent [electron0] node [left=1pt] {$C$}
        }
                     edge from parent [electron0] 
        node [left=2pt] {$A$}
    }
};
\end{tikzpicture}
\caption
[Figure 3]
{\textbf{Allowed} -- no repeated homogeneous nodes.}
\end{figure}

\begin{figure}[h]
\vspace{1mm}
\centering
\begin{tikzpicture}[thick]
\coordinate
    child[grow=down]{
        edge from parent [electron0] node [left=1pt] {$A$}
    }
    child[grow=up, level distance=0pt] {
   child  {
        child   [grow=north] {
            edge from parent [electron0] node [right=1pt] {$B$}
        }
          child [grow=south east]{ 
            edge from parent [electron0] node [right=2pt] {$C$}
        }
       edge from parent [electron0]
        node [right=1pt] {$A$}
    }
    child {
         child [grow=south west]{ 
            edge from parent [electron0] node [left=2pt] {$B$}
        }
       child  [grow=north]{ 
            edge from parent [electron0] node [left=1pt] {$C$}
        }
                     edge from parent [electron0] 
        node [left=2pt] {$A$}
    }
};
\end{tikzpicture}\hspace{8mm}
\raisebox{-4mm}{\begin{tikzpicture}[thick]
\coordinate
    child[grow=down, level distance=95pt]{
     node {{\small \emph{`Past'}}}
        edge from parent [arr0] 
    }
        child   [grow=north] {
            node {{\small \emph{`Future'}}}
            edge from parent [arr1] 
        };
\end{tikzpicture}}
\caption
[Figure 4]
{Adding a `time axis'.}
\end{figure}
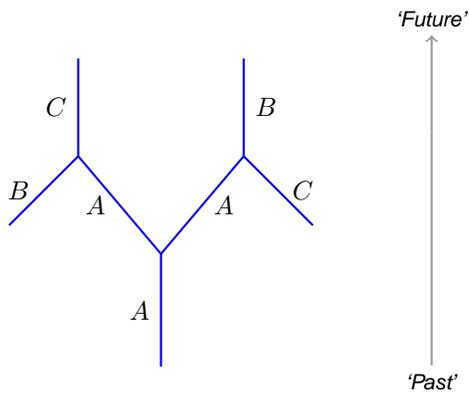

\clearpage

\begin{figure}[h]
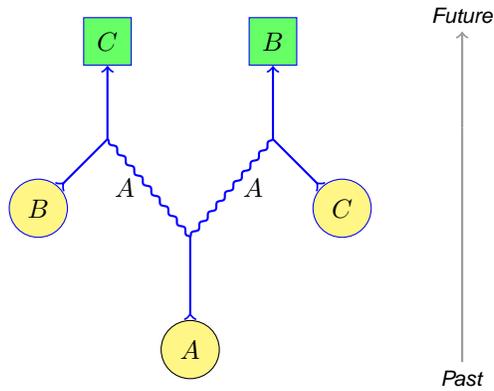

\vspace{3mm}
\centering
\helABCa\hspace{6mm}
\raisebox{-2mm}{\timearrow}
\vspace{3mm}
\caption
[Figure 5]
{Adding `preparations' and `observations'.}
\end{figure}


For future reference, let's also emphasise that the direction of causation has been `put in by hand', in this model, by our stipulation of what we can control. (It is certainly isn't given to us by the basic rules!) Our next tasks are (i) to explain what retrocausality amounts to, when the direction of causation is simply put in by hand in this way; and (ii)~to show that the model requires retrocausality.

\section[Reverse causation v. retrocausation]{Reverse causation \emph{v.} retrocausation}

Since the direction of causation is put in by hand, we could put it in `backwards', as in Fig.~6. Call this \emph{reverse} causation:  it corresponds to what causation looks like from the point of view of a creature whose time-sense is the reverse of our own. Since the 
\begin{figure}[b]
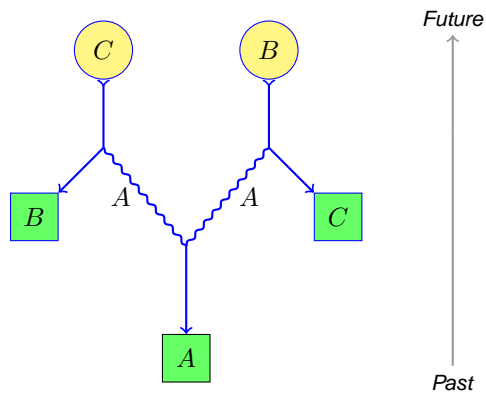

\vspace{0mm}
\centering
\helsinkz{C}{B}{B}{A}{A}{C}{A}\hspace{6mm}
\raisebox{-2mm}{\timearrow}
\vspace{3mm}
\caption
[Figure 6]
{\textbf{Reverse} causation -- interventions `from the future', observations `to the past'.}
\end{figure}
\begin{figure}[ht]
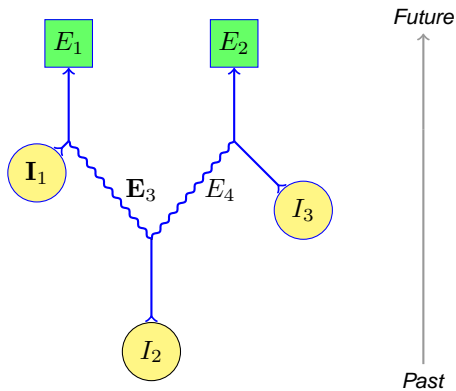

\vspace{3mm}
\centering
\helsinkRETRO{{E}_1}{E_2}{\mathbf{I}_1}{\mathbf{E}_3}{E_4}{I_3}{I_2}\hspace{6mm}
\raisebox{-2mm}{\timearrow}
\vspace{3mm}
\caption
[Figure 7]
{\textbf{Retro}causation -- an intervention $\mathbf{I}_1$  may act `backwards' on $\mathbf{E}_3$.}
\end{figure}
Helsinki model is trivially time-symmetric (with the temporal interpretation we've given it), and the causal arrow is simply put in by hand, it is no big surprise to learn that it could be put in with the opposite orientation. And if this is what retrocausality amounted to, it could hardly be big news, surely, if we applied it to quantum mechanics?


Quite so, and the interesting case is entirely different. (One of the virtues of the Helsinki model is that it displays the difference so clearly.) The interesting case is when ordinary interventions (`from the past')  make a difference \emph{prior to the intervention} -- e.g., in the notation of Fig.~7,  if the choice of the  `measurement setting' $\mathbf{I}_1$ affects the `hidden state' $\mathbf{E}_3$. (Here, think of the `$I_i$' and `$E_j$' as variables, representing the values of the three inputs, or \emph{Interventions,} and some (potential) \emph{Effects,} respectively. Each variable is restricted to the three values $A$, $B$ or $C$, of course, by the rules of the model. The position of the input node labelled `$\mathbf{I}_1$' is intended to indicate that the choice of the value of $\mathbf{I}_1$ can be made \emph{after} the time of the central 'pair production'.) This kind of influence -- when the choice of $\mathbf{I}_1$ makes a difference to $\mathbf{E}_3$ -- is what I want to call \emph{retro}causation.

Unlike  the case of reverse causation, which we can simply put in by hand -- just a different choice of hand, so to speak -- it is far from obvious that the Helsinki model involves retrocausation. To show that it does in fact do so, we need to investigate the patterns of correlations between inputs and hidden states allowed by the rules of the model. What we are looking for is a case in which a change in the left or right-hand input variables \emph{requires} a change in the hidden state.

\section{Retrocausality in Helsinki}

\newcommand\intervene[3]{$#1$\raisebox{-1mm}{$#2$}$#3$}
\newcommand\intervenee[3]{{$#1\hspace{-2.5pt}\raisebox{-1mm}{$#2$}\hspace{-0.5pt}#3$}}
\newcommand\interveneeee[3]{{$#1\hspace{-1.5pt}\raisebox{-1mm}{$#2$}\hspace{-0.5pt}#3$}}
\newcommand\interveneee[3]{{\small ${#1_{\normalsize #2}}#3$}}

To reveal the retrocausality in the Helsinki model, let's first consider 
 the admissible three-input interactions (as in Fig.~5, for example). Exploiting the obvious symmetries of the model, there are effectively only four different choices of the three inputs. Writing the choice of inputs shown in Fig.~5 as `\intervenee{B}{A}{C}', for example, the four possibilities are \intervenee{A}{A}{A}, \interveneeee{A}{A}{B}, \intervenee{B}{A}{B}, and \intervenee{B}{A}{C} itself. For each of these choices of inputs, we want to know which of the nine possible hidden states -- i.e., $\langle AA\rangle$, $\langle AB\rangle$, $\langle AC\rangle$, $\langle BA\rangle$, $\langle BB\rangle$, $\langle BC\rangle$, $\langle CA\rangle$, $\langle CB\rangle$ and $\langle CC\rangle$ -- are compatible with that choice. The fact that we have restricted ourselves to the case in which the central input is $A$ immediately excludes most of these hidden states: the only admissible possibilities are $\langle AA\rangle$, $\langle BC\rangle$ and $\langle CB\rangle$. (The notation `$\langle XY\rangle$' is intended to indicate that $X$ is the flavour of the hidden edge on the left, and $Y$ that of the hidden edge on the right.)
 
 This gives us only twelve cases to consider -- four choices of inputs, and three hidden states for each -- and the results are summarised in the State Table in Fig.~8. Note in particular that  the inputs \interveneeee{A}{A}{A} and \interveneeee{A}{A}{B} \emph{exclude} the hidden state $\langle AA\rangle$. This is the key to the model's retrocausality.

\begin{figure}[h]
\vspace{3mm}
\centering

\begin{tabular}{| r || c | c | c |}

  \hline
{} &$\langle AA\rangle$ & $\langle BC\rangle$ & $\langle CB\rangle$ \\ \hline\hline
   \interveneeee{A}{A}{A} & \textcolor{red}{\ding{55}} &  \ding{51} &  \ding{51}\rule[-1.5mm]{0.0mm}{0.5cm} \\ \hline
   \interveneeee{A}{A}{B} &  \textcolor{red}{\ding{55}} &  \ding{51} &  \ding{51}\rule[-1.5mm]{0.0mm}{0.5cm} \\ \hline
   \intervenee{B}{A}{B} & \ding{51} &  \ding{51} &  \ding{51}\rule[-1.5mm]{0.0mm}{0.5cm} \\ \hline
     \intervenee{B}{A}{C} & \ding{51}  & \ding{51}  & \ding{51}\rule[-1.5mm]{0.0mm}{0.5cm}  \\ \hline
  \end{tabular}
\vspace{2mm}
\caption
[Figure 8]
{The State Table.}
\end{figure}

\subsection{Retrocausality revealed}

Consider the case shown in Fig.~9. If either of the `measurement settings' (i.e., the left or right inputs) were an $A$, as in Fig.~10,  then the hidden state \emph{couldn't be} $\langle AA\rangle$. (In the case shown in Fig.~10, with input $A$ on the right, the two possibilities are a hidden state $\langle BC\rangle$ with left and right outputs both $B$; or a hidden state $\langle CB\rangle$ with left and right outputs  $A$ and $C$.)
So in any actual case of the kind shown in Fig.~9, the hidden state depends \emph{retrocausally} on the fact that neither `observer' chose to input the measurement setting $A$ rather than the measurement setting $B$. (As in Fig.~7, we could easily vary the position of the input nodes, to make it clear that the choice of measurement setting does not have to be made until \emph{after} the pair production that produces the hidden state.) 

Note also -- comparing Fig.~9 and Fig.~10 -- that the \emph{output} on the left depends on the measurement setting on the right. If the actual case is as shown in Fig.~9, then again we have a counterfactual dependency, apparently: if we \emph{had} chosen the input $A$ on the right, we would have obtained either the output $B$ or the output $A$ on the left, rather than the output $A$. So we also have a kind of \emph{nonlocality.}

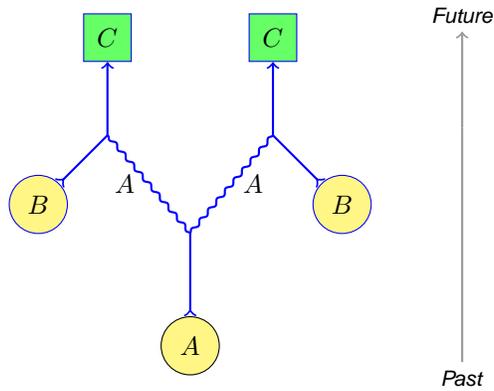
\begin{figure}[h]
\vspace{3mm}
\centering
\helABBa\hspace{6mm}
\raisebox{-2.5mm}{\begin{tikzpicture}[thick]
\coordinate
    child[grow=down, level distance=95pt]{
     node {{\small \emph{Past}}}
        edge from parent [arr0] 
    }
        child   [grow=north] {
            node {{\small \emph{Future}}}
            edge from parent [arr1] 
        };
\end{tikzpicture}}
\vspace{3mm}
\caption
[Figure 9]
{Hidden state  $\langle AA\rangle$ is possible with these inputs.}
\end{figure}

\begin{figure}[h]
\vspace{3mm}
\centering
\helsink{B|A}{B|C}{B}{{\footnotesize B|C}}{{\footnotesize C|B}}{{\textbf{A}}}{A}
\hspace{6mm}
\raisebox{-3mm}{\begin{tikzpicture}[thick]
\coordinate
    child[grow=down, level distance=95pt]{
     node {{\small \emph{Past}}}
        edge from parent [arr0] 
    }
        child   [grow=north] {
            node {{\small \emph{Future}}}
            edge from parent [arr1] 
        };
\end{tikzpicture}}
\vspace{3mm}
\caption
[Figure 10]
{A different hidden state is enforced by the change of right input.}
\end{figure}
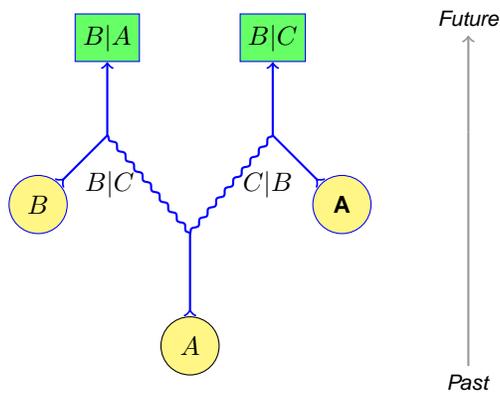

\section{Is the model consistent?}

At this point there are some questions we might raise about the consistency of the model as a whole. 
For one thing, we might wonder whether are there larger systems constructed according to the same rules in which some choice of inputs allows \emph{no} consistent assignment of outputs?  The answer to this question seems to be `no'. For suppose the contrary, and let $N$ be the minimum length for the maximal string of sequential nodes in such an inconsistent structure. If $N>1$, we could obtain a shorter inconsistent structure by choosing an inconsistent structure of length $N$, removing its lower-most level, and supplying by hand to the next level the inputs otherwise supplied by lower-most level. But this would contradict the assumption that $N$ is the minimum length for such a structure, so $N=1$. But there is no such system of length $1$, apparently, and so no system of greater length, either.

This consistency property means that in interpreting the model in terms of our intuitive ideas of intervention, control and observation, we don't need to impose any restrictions on the ` free choices' of our toy physicists, in order to preserve consistency. This is an interesting result, especially in the light of the fact that two of the standard concerns about retrocausal models are that they might conflict with free will, and/or lead to inconsistencies or paradoxes of some kind.

\subsection{Causal loops}

We can take this concern with consistency under the natural interpretation a stage further, by allowing our toy physicists a freedom real physicists have in the case of EPR/Bell experiments, namely, to perform the two measurements at sufficiently different times, so that the result of one can be allowed to influence the setting of the other (by an ordinary `classical' causal channel). This possibility is a recognised source of potential causal loops, in retrocausal models of EPR/Bell situations; see, e.g., (Berkovitz 2002).

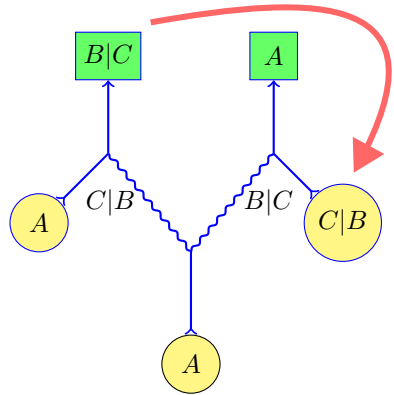
\begin{figure}[ht]
\vspace{3mm}
\centering\mbox{\helsink{B|C}{A}{A}{C|B}{B|C}{C|B}{A}\hspace{-32mm}
\raisebox{29mm}{\begin{tikzpicture}
\draw[-triangle 60,line width=.08cm,color=red!60] (2.0,9.6) ..
controls
           (4.25,10) and (6,10) .. (4.7,7.6);
  \end{tikzpicture}}}
\vspace{3mm}
\caption
[Figure 11]
{A causal loop? Left output controls right input.}
\end{figure}

Let's represent this possibility by adding to our diagrams the kind of causal link represented by the red arrow in Fig.~11. Keep in mind that despite the way it is depicted in Fig.~11, this is not to be thought of as a \emph{retrocausal} influence. (Imagine the diagram elongated on the right, so that the right input actually occurs \emph{after} the left output.) Keep in mind also that in this version, the new causal link lies in the classical realm of our toy physicists -- it isn't part of the model itself. (As the model stands, the main obstacle to incorporating it within the model is the requirement that the two kinds of node must alternate -- otherwise, we could simply make the output of the pair annihilation on the left an input of the pair annihilation on the right, eliminating the `external' red arrow altogether.)

In the case shown in  Fig.~11, the output  $B$ on the left produces input $C$ on the right, and the output  $C$ on the left produces input $B$ on the right. There is a consistent assignment of hidden states and 
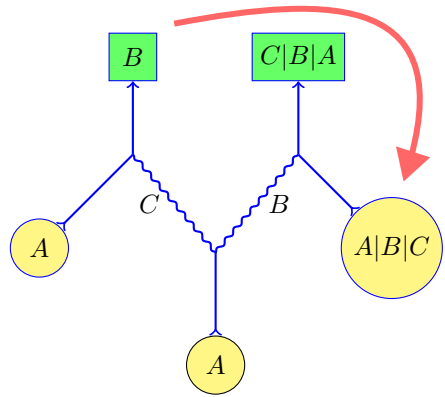
\begin{figure}[ht]
\vspace{3mm}
\centering\mbox{\helsinkBIG{B}{C|B|A}{A}{$C$}{$B$}{A|B|C}{A}\hspace{-37mm}
\raisebox{28mm}{\begin{tikzpicture}
 \draw[-triangle 60,line width=.08cm,color=red!60] (2.15,9.6) ..
controls
           (4.25,10) and (6,10) .. (5.2,7.5);
  \end{tikzpicture}}}\vspace{3mm}
\caption
[Figure 12]
{Generalising the previous case.}
\end{figure}
right output in either case, showing that the constraint admits two consistent solutions (with the given choice of left and centre inputs -- i.e., $A$ in both positions).

Generalising this case, consider the three possible ways in which a left output $B$  can fix a right input, as in Fig.~12. Again, all three cases allow a consistent assignment of the right output. Exploiting the symmetries of the model once more, this is sufficient to demonstrate that  \emph{whenever} the left and centre inputs are the same, \emph{any} set of left-output-to-right-input constraints allows at least one consistent assignment of hidden states and right output --  i.e., no such constraint can `shut the system down'.

\begin{figure}[t]
\vspace{3mm}
\centering\mbox{\helsinkBIG{A}{C|B|A}{B}{$C$}{$B$}{A|B|C}{A}\hspace{-37mm}
\raisebox{28mm}{\begin{tikzpicture}
 \draw[-triangle 60,line width=.08cm,color=red!60] (2.15,9.6) ..
controls
           (4.25,10) and (6,10) .. (5.2,7.5);
  \end{tikzpicture}}}\vspace{3mm}
\caption
[Figure 13]
{A new case, with different inputs in left and centre positions.}
\end{figure}
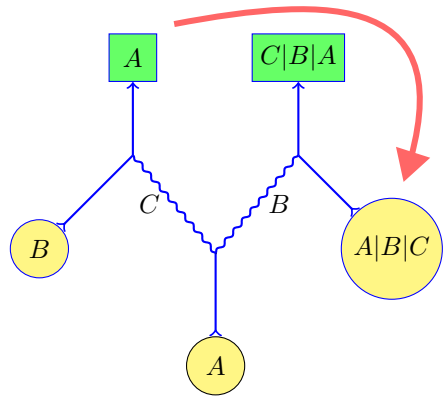

This leaves the cases in which the initial and left inputs are different. 
Here, consider, for example, the three possible ways in which a left output \emph{A} can fix a right input, as in Fig.~13.
 Again, all three possibilities allow a consistent right output.
 And again, \emph{any} set of left-output-to-right-input constraints allows at least one consistent assignment of hidden states and right output --  again, no such constraint can shut the system down.

This kind of constraint is non-trivial, however. Fig.~14 shows a case in which its effect is to exclude a hidden state --  $\langle AA\rangle$ --  that would otherwise be permitted. So the Helsinki model is rich enough to show how this kind of causal loop can impose new constraints, without leading to inconsistency.

\begin{figure}[h]
\vspace{3mm}
\centering\mbox{\helsinkBIG{C}{?}{B}{$A$}{$A$}{A}{A}\hspace{-32mm}\raisebox{26mm}{\begin{tikzpicture}
 \draw[-triangle 60,line width=.08cm,color=red!60] (2.22,9.6) ..
controls
           (4.25,10) and (6,10) .. (5.2,7.3);
  \end{tikzpicture}}}\vspace{3mm}
\caption
[Figure 14]
{A substantial constraint.}
\end{figure}
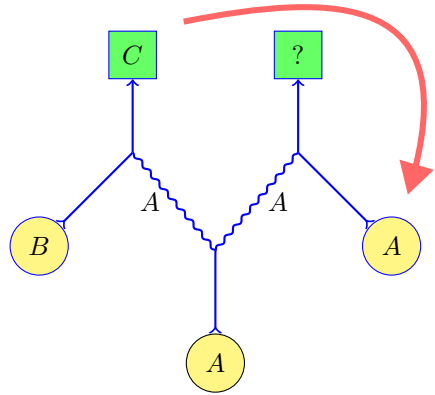

\section{Improving the model}

I conclude with a wish-list of enhancements -- further steps it would be interesting to be able to take in future iterations of the Helsinki model, or in something like it:
\begin{enumerate}
 \item Adding probabilities, and showing that in virtue of the retrocausality, they are bound to have some of the characteristics of the probabilities associated with QM amplitudes --  e.g., that probabilities of results of measurements cannot generally be regarded as probabilities of pre-existing states, regarded as \emph{independent of the choice of future measurements.}
 \item Developing an analogy between the standard QM state function and what we know in the Helsinki model \emph{if we don't know the future measurement settings} -- in other words, an analogy with the kind of epistemic `coarse graining' of the Helsinki model which would be necessary to represent the state of knowledge of a physicist (or toy physicist) who wants to make predictions with respect to a range of possible `next measurements'.
 \item Hence connecting the Helsinki model, or some variant of it, to Spekkens' `epistemic' toy models.
 
 \item Investigating the nonlocality of Helsinki-like models, in search  of the retrocausal toy modeller's Holy Grail: a model with Bell-like correlations without signalling. 
\end{enumerate}
I don't know to what extent such enhancements are possible, but I'll be pleased if the model inspires anyone to try to find out.

\section*{Abstract}
A number of writers have been attracted to the idea that some of the peculiarities of quantum theory might be manifestations of `backward' or `retro' causality, underlying the quantum description. This idea has been explored in the literature in two main ways: firstly  in a variety of explicit models of quantum systems, and secondly at a conceptual level. This note introduces a third approach, intended to complement the other two. It describes a simple toy model, which, under a natural interpretation, shows how retrocausality can emerge from simple global constraints. The model is also useful in permitting a clear distinction between the kind of retrocausality likely to be of interest in QM, and a different kind of reverse causality, with which it is liable to be confused. The model is proposed in the hope that future elaborations  might throw light on the potential of retrocausality to account for quantum phenomena.

\section*{Bibliography}

Berkovitz, J. 2002: On Causal Loops in the Quantum Realm. In T.~Placek and J.~Butterfield (eds.), \emph{Non-locality and Modality,} Kluwer, 235--257. \vspace{3pt}

\noindent Costa de Beauregard, O. 1953: M\'echanique Quantique. \emph{Comptes Rendus Acad\'emie des 
Sciences} 236, 1632. \vspace{3pt}

\noindent Costa de Beauregard, O. 2005: Personal communication to Guido Bacciagaluppi and Huw Price,  Bourron-Marlotte, France, 28.03.05.\vspace{3pt}

\noindent Price, H. 1996: \emph{Time's Arrow and Archimedes' Point.} New York: Oxford University Press.\vspace{3pt}

\noindent Spekkens, R. W. 2004: In Defense of the Epistemic View of Quantum States: a Toy Theory. \href{http://arXiv.org/abs/quant-ph/0401052}{\url{http://arXiv.org/abs/quant-ph/0401052}} \vspace{3pt}

\noindent Sutherland, R. 2006: Causally Symmetric Bohm Model. Forthcoming in \emph{Studies in the History and Philosophy of Modern Physics,} \textbf{39}, 2008. Preprint available online here: \href{http://arxiv.org/abs/quant-ph/0601095v2}{\url{http://arxiv.org/abs/quant-ph/0601095v2}}\vspace{3pt}

\noindent Wharton, K. 2007: A Novel Interpretation of the Klein-Gordon Equation. Preprint available here: \href{http://arxiv.org/abs/0706.4075}{\url{http://arxiv.org/abs/0706.4075}}\vspace{3pt}

\noindent Wiseman, H. 2005: From Einstein's Theorem to Bell's Theorem: A History of Quantum Nonlocality. \href{http://arxiv.org/abs/quant-ph/0509061v3}{\url{http://arxiv.org/abs/quant-ph/0509061v3}}

\end{document}